\documentclass{iau}
\usepackage{graphicx}

\title[Prominence cavities] %% give here short title %%
{Magnetism and the Invisible Man:  The mysteries of coronal cavities}

\author[Sarah Gibson]   %% give here short author list %%
{Sarah Gibson$^1$
 \thanks{NCAR is supported by the National Science Foundation }
 }

\affiliation{$^1$High Altitude Observatory/National Center for Atmospheric Research \\ 3080 Center Green Dr.
Boulder, CO, 80027, USA \\ email: {\tt sgibson@ucar.edu} 
}

\pubyear{2013}
\volume{300}  %% insert here IAU Symposium No.
\doi{10.1017/S1743921313010879}
\pagerange{139-146}
\jname{Nature of Prominences and their Role in Space Weather}
\editors{B. Schmieder \& J. M. Malherbe, eds.}
\begin{document}

\maketitle

\begin{abstract}
Magnetism defines the complex and dynamic solar corona.  Twists and tangles in coronal magnetic fields build up energy and ultimately erupt, hurling plasma into interplanetary space.  These coronal mass ejections (CMEs) are transient riders on the ever-outflowing solar wind, which itself possesses a three-dimensional morphology shaped by the global coronal magnetic field.  Coronal magnetism is thus at the heart of any understanding of the origins of space weather at the Earth.   However, we have historically been limited by the difficulty of directly measuring the magnetic fields of the corona, and have turned to observations of coronal plasma to trace out magnetic structure.  This approach is complicated by the fact that plasma temperatures and densities vary among coronal magnetic structures, so that looking at any one wavelength of light only shows part of the picture.  In fact, in some regimes it is the lack of plasma that is a significant indicator of the magnetic field.  Such a case is the coronal cavity: a dark, elliptical region in which strong and twisted magnetism dwells.  I will elucidate these enigmatic features by presenting observations of coronal cavities in multiple wavelengths and from a variety of observing vantages, including unprecedented coronal magnetic field measurements now being obtained by the Coronal Multichannel Polarimeter (CoMP).  These observations demonstrate the presence of twisted magnetic fields within cavities, and also provide clues to how and why cavities ultimately erupt as CMEs. 
\end{abstract}

\firstsection % if your document starts with a section,
              
              \vspace{5mm}
              
              {\it ``No hand -- just an empty sleeve... Then, I thought, there's something odd in that.  What the devil keeps that sleeve up and open, if there's nothing in it?"} -- The Invisible Man (\cite[Wells 1897]{wells_97})
              
\section{Introduction}

%Why does the number of sunspots wax and wane with a cycle of approximately 11 years?
%What heats the outer atmosphere of the sun, the solar corona, to a million degrees, above a star whose surface temperature is less than 6000?
%What accelerates the solar wind to super-sonic speeds?
%And, finally, what causes a billion tons of coronal plasma to be hurled out into interplanetary space at a million miles per hour?
%The answers to all these outstanding mysteries of solar physics lies in the Sun's magnetic fields.
%In this review I will talk about magnetism in the Sun's corona, and the clues we can gain from observations -- sometimes even when there appears to be nothing there.

Coronal cavities are dark, elliptical structures that surround prominences (Figure 1).  Like prominences, cavities are long-lived and may be stable for days or even weeks (\cite[Gibson {\it et al.} 2006]{gibcav}).  Also like prominences, cavities exhibit dynamic behavior even when not erupting, with swirling flows of coronal plasma within the cavity tracing out helical structure (\cite[Li {\it et al.} 2012]{li_12}).   Cavities do eventually erupt along with their embedded prominences as coronal mass ejections (CMEs): roughly a third were observed to do so in a survey of over one hundred polar-crown-filament (PCF) cavities (\cite[Forland {\it et al.} 2013]{forland_13}).  Since a typical (median) length of time these cavities were visible at the limbs was about four and a half days, the time spent at the two limbs during the approximately 27-day solar rotation is about one-third; this implies that if one could observe all the way around the Sun, one would see all the cavities erupt eventually.

Since cavities represent the bulk of the combined erupting prominence-cavity volume, it is their magnetic structure that maps to the magnetic cloud passing the Earth.  If cavities are the ``Invisible Man'' of solar physics, prominences are his footprints: more visible perhaps, but only representative of a fraction of the magnetic structure that erupts.  Cavities thus are key to understanding the nature of pre-CME equilibria and the mechanisms that trigger their loss.

But how does one measure the invisible? Luckily, as I will describe below, cavities are not truly empty.  Their detection is subject to stringent line-of-sight constraints, and it is likely that many remain unobserved because of obscuration by surrounding bright structures in the optically-thin corona.  However, recent work modeling the 3D geometry of PCF cavities, combined with new observations, has enabled a detailed analysis of cavity physical properties.  A self-consistent picture has emerged explaining cavity morphology, sub-structure, and dynamic evolution of the cavity that is consistent with the theory that the cavity is a magnetic flux rope (e.g., \cite[Low \& Hundhausen (1995)]{lowhund}).

In Section 2, I will review observations of stable (non-erupting) cavity plasma and magnetic properties.  In Section 3, I will present present observations of cavities in relation to CMEs.  In Section 4, I will conclude by discussing how these observations map to magnetic flux ropes.

\begin{figure}[ht]
\begin{center}
 \includegraphics[width=5in]{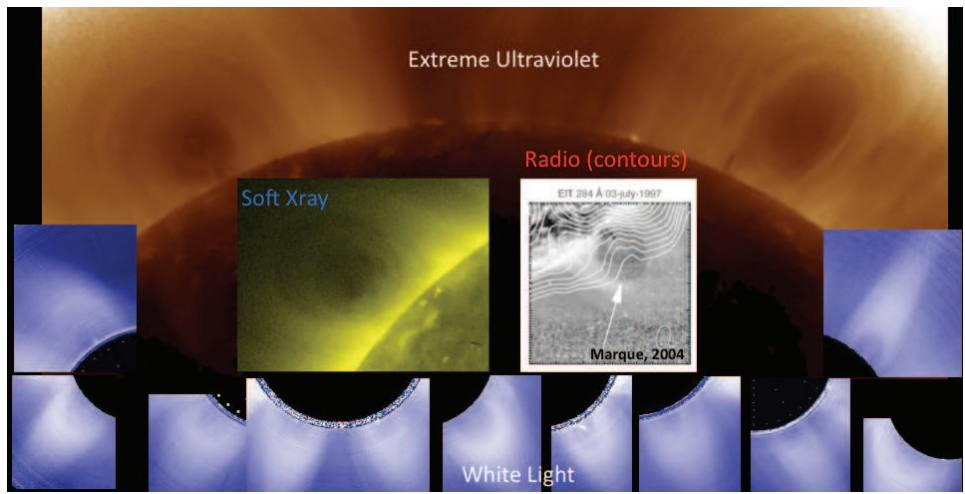} 
 \caption{Cavities are visible in a broad range of wavelengths.  Extreme ultraviolet (EUV) observations from Solar Dynamics Observatory/Atmospheric Imaging Assembly (SDO/AIA); Soft-Xray (SXR) from Hinode X-ray Telescope (XRT); white light images from Mauna Loa Solar Observatory Mk4 K-coronameter (MLSO/Mk4); Radio contours (Nancay) overlaid on Solar and Heliospheric Observatory EUV Imaging  Telescope (SOHO/EIT) observations (\cite[Marque 2004]{marque_04}).}
\end{center}
\end{figure}
\vspace{-.5cm}

\section{Coronal cavities: Observations}

Cavities were first observed in white light in eclipses (see e.g. \cite[Waldmeier (1970)]{waldmeier_70}, \cite[Tandberg-Hanssen (1974)]{tandberg_74}). The advent of coronagraphs, radio, EUV and SXR telescopes have given us a means to observe them on a daily basis  (Figure 1).    In a six-year study of white light images, \cite[Gibson {\it et al.} (2006)]{gibcav} found 98 distinct cavity systems, with cavities visible approximately one in ten days (Figure 2, left).  However, these observations were taken with an occulted coronameter, so only cavities with heights $> 1.15$ solar radius could be identified.  In a 19-month-long study of cavities at EUV wavelengths, \cite[Forland {\it et al.} (2013)]{forland_13} found 129 distinct cavity systems, with cavities visible $78\%$ of the days (Figure 2, right).  This survey was able to measure many smaller cavities than would have been occulted in the white light survey.  Another difference between the surveys was that, while the white light survey encompassed years of solar maximum, when a complexity of bright structures along the line of sight may well have obscured cavities, the EUV survey took place during the ascending phase of the solar cycle, a time when PCFs were common.  PCFs are large, longitudinally-extended, quiescent filaments at high latitudes, and as such present near-ideal viewing conditions for cavities. 

\begin{figure}[ht]
\begin{center}
 \includegraphics[width=5in]{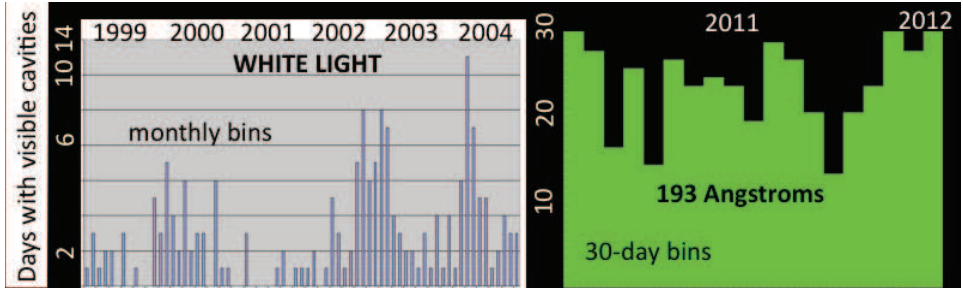} 
 \caption{Cavities are ubiquitous.  Left: white light survey of cavities from November 1998-September 2004 using MLSO/MK4 coronameter observations (\cite[Gibson {\it et al.} 2006]{gibcav}; figure courtesy Joan Burkepile).  Right: EUV (193 \AA~) survey of cavities from June 1, 2010 - Dec 31, 2012 using SDO/AIA images (\cite[Forland {\it et al.} 2013]{forland_13}).}
\end{center}
\end{figure}

\cite[Gibson {\it et al.}  (2010)]{gibson_10} studied the 3D morphology of a cavity using observations at multiple wavelengths, vantage points, and covering multiple days.  The cavity was modeled as a tunnel-like structure, with a Gaussian height (Figure 3, left) and elliptical cross-section.  \cite[Forland {\it et al.} (2013)]{forland_13} fit ellipses to all the EUV cavities in the survey, and found a strong tendency ($93\%$) for cavity ellipses to be taller than they were wide.
\vspace{-1.5cm}

   \begin{figure}[ht!]
%\begin{center}
 \includegraphics[angle=270,width=4.5in]{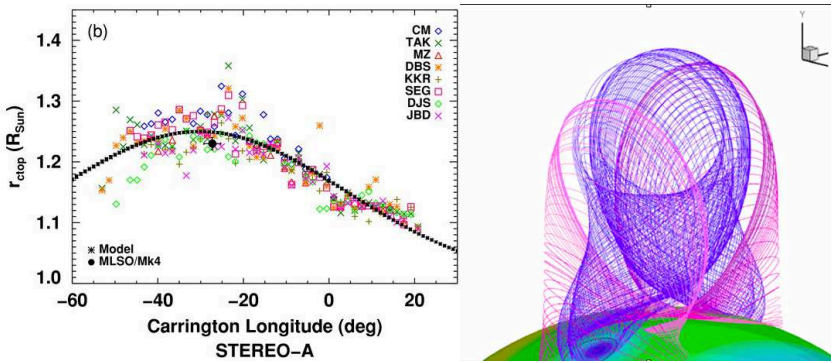}
 \vspace{-2.cm}  
 \caption{Cavities have arched, tunnel-like morphologies with elliptical cross-sections. Left: from \cite[Gibson {\it et al.} (2010)]{gibson_10}; cavity ellipse height vs. longitude/date. Right: flux surfaces of \cite[Gibson \& Fan (2006)]{gibfan_06b} simulation of flux rope.}%\end{center}
\end{figure}
  
Building on the 3D morphology found by \cite[Gibson {\it et al.}  (2010)]{gibson_10}, \cite[Schmit \& Gibson (2011)]{schmit_11} extracted density of a coronal cavity from multiwavelength observations.  The density was found to be approximately $30\%$ depleted at the center of the cavity relative to the surrounding streamer at the same height.  This was consistent with prior analyses which found that cavities were, in general, significantly more dense than coronal holes (thus, not truly ``invisible'') and possessed on average $25\%$ depletion and maximum $60\%$ depletion relative to the surrounding streamer (\cite[Fuller \& Gibson 2009]{fuller_09}). Building on both the morphology and the density analyses, \cite[Kucera {\it et al.} (2012)]{kucera_12})  found that the average temperature in the cavity was similar to that of the surrounding streamer (about $1.5 MK$); however, the cavity exhibited more thermal variability, indicating multiple temperatures were present at a given height.

\begin{figure}[ht]
\begin{center}
\vspace{-.2cm}
 \includegraphics[width=4.5in,height=2.5in]{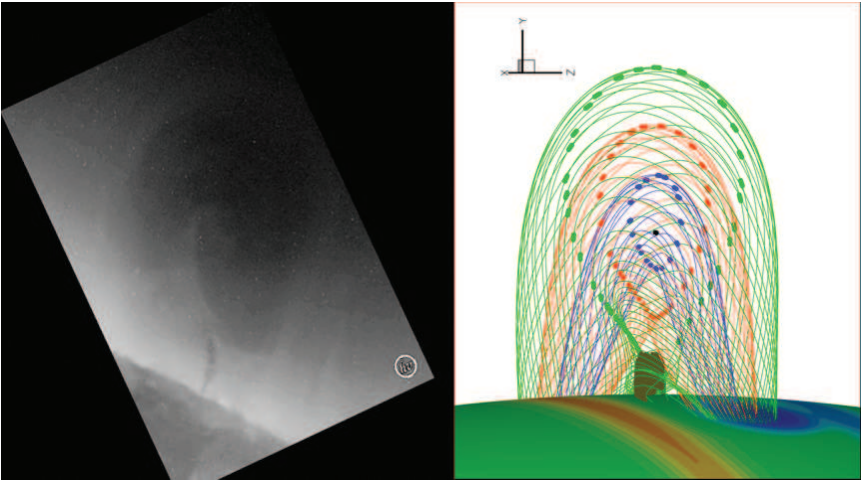} 
 \caption{Cavities have substructure.  Left: July 14, 2013 observations of a cavity from SDO/AIA (193 \AA~), showing prominence and  horn within larger-scale cavity. Right: magnetic flux surfaces (colored lines, with dots indicating intersection with plane of sky) and dips in field lines (lower sheet of brown dots) in magnetic flux rope simulation (\cite[Gibson \& Fan 2006]{gibfan_06b}).}
\end{center}
\end{figure}

\vspace{-.3cm}

The thermal variability within cavities is likely related to their often dynamic substructures. In particular, disk or ring-like structures lying at the center of the cavity (sometimes referred to as ``chewy nougats'') are commonly observed in soft X-ray, indicating regions of elevated temperature (\cite[Hudson et al. 1999]{hudson_99}; \cite[Reeves et al. 2012]{reeves_12}).   These nougats sometimes appear immediately above the prominence, like a lollypop on a stick.  In EUV, flows trace out horn-like structure in a similar central location above the prominence (Figure 4; left), and are temporally and spatially linked to flows in the cooler prominence plasma (\cite[Schmit \& Gibson 2013]{schmit_13}).  Flows along the line of sight have also been measured (\cite[Schmit {\it et al.} 2009]{schmit_09}).  These flows are of order $5-10~ km/sec$, have length scales of tens of megameters, and persist for at least one hour.  They typically have outer boundaries corresponding to that of the cavity or its central substructure, and occasionally exhibit nested ring-like structure (\cite[B{\c a}k-St{\c e}{\'s}licka {\it et al.} 2013]{ula_13}) (Figure 5; top).

\begin{figure}[ht!]
\begin{center}
 \includegraphics[angle=270,width=4.7in]{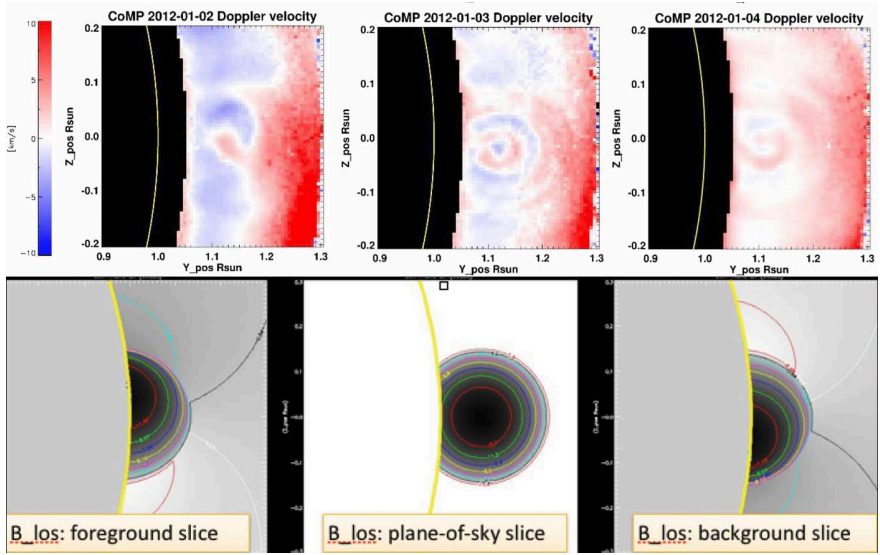} 
 \vspace{-.3cm}
 \caption{Cavities contain line-of-sight flows, sometimes with a bulls-eye pattern.  Top: MLSO Coronal Multichannel Polarimeter (CoMP) Doppler velocity observations for a cavity seen over three days (from \cite[B{\c a}k-St{\c e}{\'s}licka {\it et al.} (2013)]{ula_13}).  Bottom:  line-of-sight magnetic field in the plane of sky for flux rope model of \cite[Low \& Hundhausen (1995)]{lowhund}. }
\end{center}
\end{figure}

Recently, a new means of directly probing the magnetic structure of cavities has become available through the Coronal Multichannel Polarimeter (CoMP):  a coronagraph that measures Stokes polarimetry vectors and line-of-sight velocities using optically-thin coronal emission lines (\cite[Tomczyk {\it et al.} 2008]{tomczyk_08}).  Linear polarization ($\sqrt{Q^2+U^2}$, where $Q$ and $U$ are Stokes vectors) has turned out to be a particularly useful diagnostic for coronal cavities
%By using forward-modeling techniques, 
(\cite[Dove {\it et al.} 2011]{dove_11}).
% demonstrated that the ring of linearly-polarized light observed by CoMP within a cavity matched that predicted by a spheromak-type flux rope (e.g., \cite[Gibson \& Low (1998)]{giblow_98}).  This observation was taken in 2005 before CoMP was deployed to MLSO in Hawaii, and was for a large, but not PCF cavity.  
Over the past few years, CoMP has shown that the linear polarization of PCF cavities systematically exhibit a structure akin to that of a rabbit's head (``lagomorph'')(Figure 7; 
\cite[B{\c a}k-St{\c e}{\'s}licka {\it et al.} (2013)]{ula_13}).
  As is evident by comparing Figures 6 and 7, linear-polarization lagomorphs generally scale with cavity size (\cite[B{\c a}k-St{\c e}{\'s}licka {\it et al.} 2014 (this issue)]{ula_14}).

\begin{figure}[ht]
\begin{center}
 \includegraphics[width=4.7in]{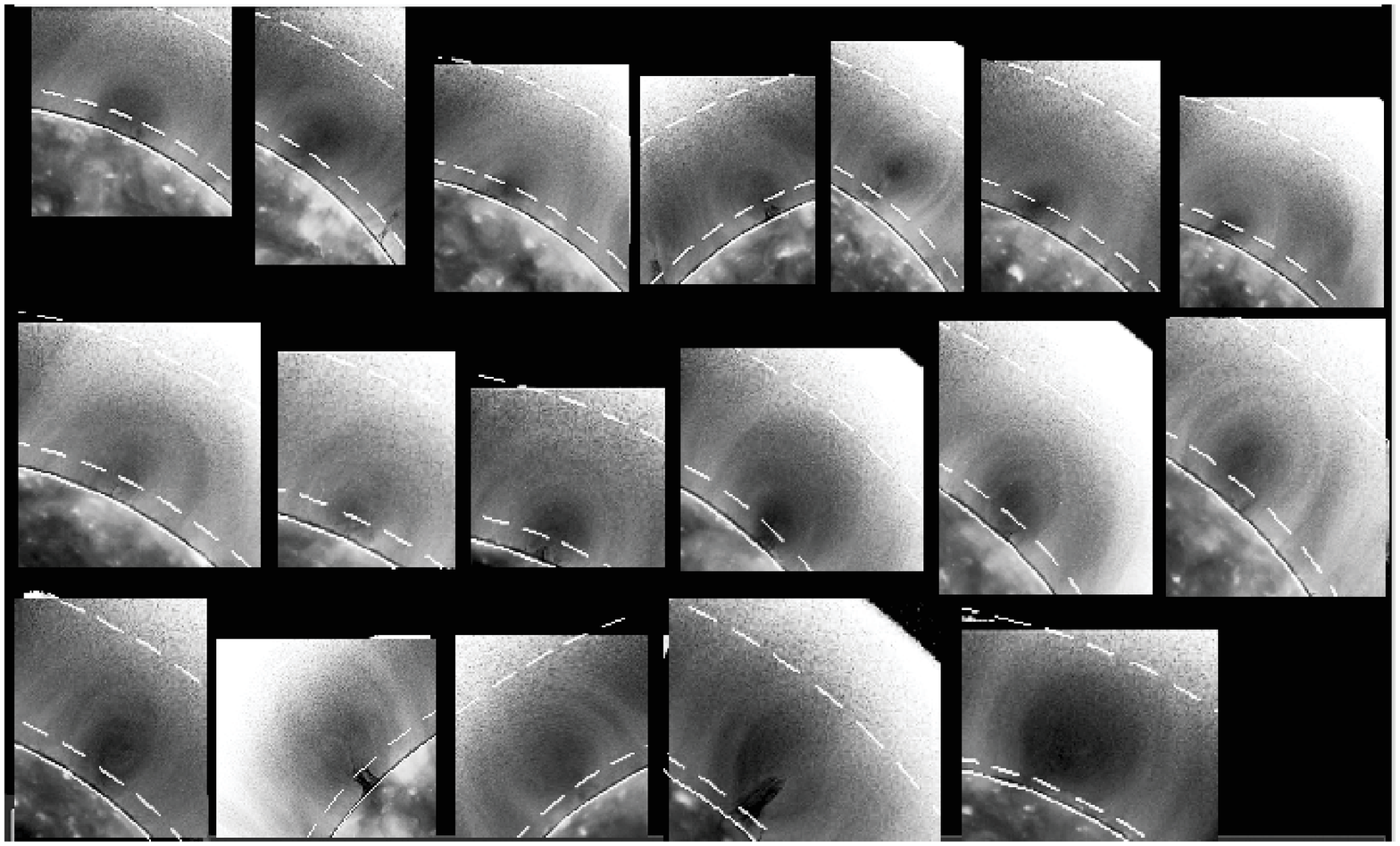} 
 \caption{Cavities of a variety of sizes and shapes as seen in SDO/AIA 193 \AA. Solid white line is at solar photosphere; dashed white line is at 1.05 solar radii (location of occulting disk for MLSO/CoMP telescope).  Dates, starting at upper left:  5/25/11; 6/14/11;
6/24/11;
7/9/11;
7/14/11;
7/26/11;
7/27/11.  
Next row:
7/28/11;
7/29/11;
8/1/11;
8/10/11;
8/11/11;
8/12/11;
Next row:
8/14/11;
8/24/11;
8/30/11;
11/11/11;
1/2/12.}
\end{center}
\end{figure}

\begin{figure}[ht!]
\vspace{-1cm}
\begin{center}
 \includegraphics[width=4.7in]{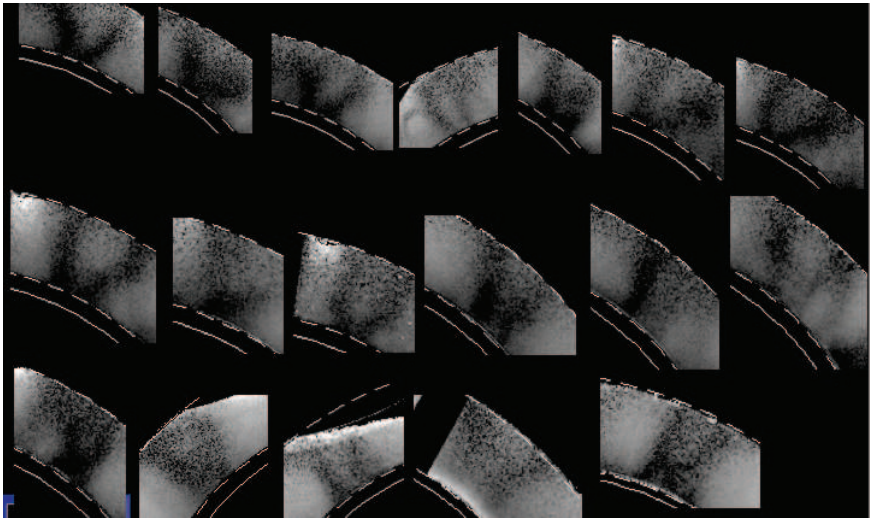} 
 \vspace{-1cm}
 \caption{Linear polarization lagomorphs corresponding to the cavities of Figure 6, as seen by MLSO/CoMP.}\end{center}
\end{figure}

\section{Cavites and CMEs}

%Cavities �bodily� erupt as CMEs. (\cite[Gibson {\it et al.} (2006)]{gibcav}, \cite[Mari\v{c}i\'{c} {\it et al.} (2009)]{maricic_09}
  
 The properties of cavities prior to eruptions and their evolution leading up to CMEs may provide clues to the mechanisms that trigger them.  
 \cite[Gibson {\it et al.} (2006)]{gibcav} found an upper limit to cavity height of approximately $1.5$ solar radii (Figure 8).  This may imply a global limit beyond which cavities are unstable.  The EUV cavities of the survey of \cite[Forland {\it et al.} (2013)]{forland_13} lie well below this height, but Figure 8 (right) indicates a slight tendency for  higher cavities to be eruptive (red diamonds) rather than not (green triangles).
  
 Perhaps the strongest indicator of an impending eruption is the shape of the cavity.  As seen in Figure 8, cavities tend to have aspect ratios less than one (i.e., their width is smaller than their height).  Moreover, eruptive cavities in general have smaller aspect ratios than non-eruptive cavities.  These aspect ratios are based on fitting the cavities with elliptical shapes.  \cite[Forland {\it et al.} (2013)]{forland_13}  noted that in some cases cavities were better characterized as teardrop-shaped, and found  that 68\% of teardrop-shaped cavities erupted as CMEs as compared to 23\% of elliptical cavities (and 10\% of semicircular cavities). 
\cite[Gibson {\it et al.} (2006)]{gibcav} noted similar behavior; due to the occulting disk, the full shape of the white-light cavities was not measured, but a quality referred to as ``necking'' was noted when cavities had narrower bases than tops.  They found that 10/10 cases of cavities which erupted within 24 hours had necking, vs 25/99 of the entire sample.

  \begin{figure}[ht!]
\begin{center}
 \includegraphics[width=5in]{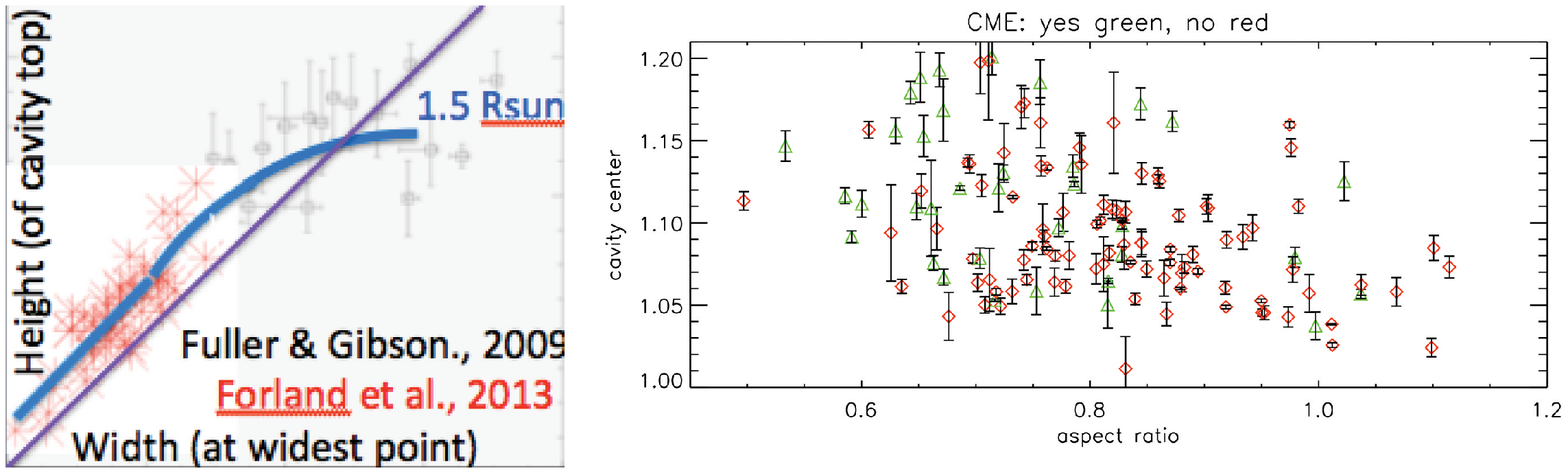}
 \caption{Cavity height and aspect ratio vs CME.  Left: the cavities of the \cite[Fuller \& Gibson (2009)]{fuller_09} white-light cavity study (error bars) and \cite[Forland {\it et al.} (2013)]{forland_13} EUV cavity survey (red asterisks).  Most cavities have aspect ratio less than one; cavity heights do not in general reach higher than about 1.5 solar radii.  Right: cavity center heights tend to be higher for those that erupt (green triangles) than those that don't (red diamonds), and cavity morphology tends to be more narrow (small aspect ratio) (\cite[Forland {\it et al.} 2013]{forland_13}).}\end{center}
\end{figure}

\section{Conclusions: cavities as magnetic flux ropes}

Cavities are observed to be ubiquitous (Figures 1 and 2).  If cavities are flux ropes, this is to be expected.  A constant-alpha force-free state is the minimum energy configuration for a given boundary condition (\cite[Woltjer 1958]{woltjer_58}): given sufficient helicity, this will be a flux rope. A large-scale force-free equilibrium of minimum energy conserving helicity can be reached through turbulent inverse cascade of helicity, from small scale to large (\cite[Taylor 1974]{taylor_74}). Since helicity is very nearly conserved even through magnetic reconnection  (\cite[Berger \& Field 1984]{bergerfield_84}), the free energy stored in the still-twisted large-scale magnetic fields represent ``flare un-releasable'' magnetic free energy (\cite[Zhang \& Low 2005]{zhanglow_05}).

Cavities have arched, tunnel-like morphology with skinny-elliptical cross-section.  Simulations have demonstrated that a flux rope expanding upwards into closed magnetic fields  may find an equilibrium configuration as the forces causing the upward expansion are countered by confining magnetic tension forces.   The equilibrium flux rope will then have an arched, tunnel-like morphology with narrow aspect ratio (Figure 3 (right)).   

Cavities have low density, substructure, and are multithermal and dynamic (Figures 4-5).
 \cite[Schmit \& Gibson (2014)]{schmit_14} (this issue) used hydrostatic models to argue that field lines at the center of the cavity, which are arched and non-dipped and relatively short (see blue lines in Figure 4), will have low density relative to surrounding winding/dipped field lines.  The degree of depletion found was about $30\%$ for a flux rope of aspect ratio (width, height, length) reasonable for a PCF.  \cite[Schmit \& Gibson (2013)]{schmit_13} argued that dynamic flow along dipped field lines  driven, for example, by thermononequilibrium (\cite[Antiochos {\it et al.} 1999]{antio_99}) provided an explanation for EUV horns above the cavity consistent with observations of cavities and prominences.   Alternatively, \cite[Fan (2012)]{fan_12} argued that heating and reconnection-driven flows at the top of the prominences could explain these structures, as well as the elevated temperatures of chewy nougats.  In general, flows along magnetic flux surfaces, particularly those where dynamics might be expected such as at the interface of dipped and non-dipped field, may explain disk and ring-like structures and flows within cavities. Figure 5 (bottom) demonstrates this;  if flows are field-aligned and assuming constant velocity, the line-of-sight component would peak at a flux rope's axis. Moreover, the shift of the flux rope in front or behind the plane of sky might introduce asymmetries in line-of-sight flow such as have been observed.

Polar crown filament cavities exhibit lagomorphic linear polarization signals (Figures 6-7).
\cite[B{\c a}k-St{\c e}{\'s}licka {\it et al.} (2013)]{ula_13}
used forward modeling techniques to demonstrate that such lagomorphs are to be expected for a cylindrical flux rope extended along the line of sight.  \cite[Rachmeler {\it et al.} (2013)]{rachmeler_13} discussed the often subtle differences between the flux-rope and the sheared-arcade model linear polarization signatures that might be expected in PCF cavities.
Interestingly, \cite[Dove {\it et al.} (2011)]{dove_11}
found a different, ring-like linear polarization signal in a cavity that matched that predicted by a spheromak-type flux rope (e.g., \cite[Gibson \& Low (1998)]{giblow_98}).  This observation was taken in 2005 before CoMP was deployed at MLSO in Hawaii, and was for a large, but not PCF cavity.  

The clear association of CMEs with high, narrow, and teardrop-shape cavities is an intriguing clue to why eruptions occur.  Such a shape may occur if, for instance, a current sheet forms below a flux rope. If this is followed by reconnections at this current sheet and the slow rise of the flux rope, the increased height for cavities immediately prior to eruption may be explained.  This is the picture painted by simulations which find such behavior leading up to the ultimate loss of stability through a ``torus instability''  (\cite[Aulanier {\it et al.} 2010]{aulanier_10}, \cite[Savcheva {\it et al.} 2012]{Savcheva12c}, \cite[Fan 2012]{fan_12}). 

\vspace{.1cm}

{\bf Acknowledgements.}  I thank and acknowledge the members of the International Space Science Institute (ISSI) working groups on coronal cavities (2008-2010) and coronal magnetism (2013-2014).  I am particularly indebted to Urszula B{\c a}k-St{\c e}{\'s}licka, Giuliana de Toma, James Dove, Yuhong Fan, Blake Forland, Jim Fuller, Terry Kucera, B. C. Low, Laurel Rachmeler, Kathy Reeves, and Don Schmit.

%\cite{antio_99}
%\cite{wells_97}
%\cite{gibcav}
%\cite{li_12}
%\cite{forland_13}
%\cite{marque_04}
%\cite{waldmeier_70}
%\cite{tandberg_74}
%\cite{gibson_10}
%\cite{schmit_11}
%\cite{kucera_12}
%\cite{fuller_09}
%\cite{schmit_13a}
%\cite{schmit_14}
%\cite{hudson_99}
%\cite{reeves_12}
%\cite{gibfan_06b}
%\cite{schmit_09}
%\cite{lowhund}
%\cite{ula_13}
%\cite{tomczyk_08}
%\cite{dove_11}
%\cite{ula_14}
%\cite{woltjer_58}
%\cite{taylor_74}
%\cite{fan_12}
%\cite{Savcheva13c}
%\cite{giblow_98}
%\cite{aulanier_10}
%\cite{bergerfield_84}
%\cite{rachmeler_13}
%\cite{zhanglow_05}

%\bibliography{../../mybibliography}  

\vspace{-.5cm}

%\end{document}

\end{document}